\documentclass[nofootinbib,notitlepage, aps]{revtex4-1}%

\usepackage{slashed,amsmath,amssymb}
\usepackage{graphicx}
\usepackage{dcolumn}
\usepackage{bm}
\usepackage{dsfont}
\usepackage{xcolor}

\usepackage{color} \definecolor{darkgreen}{rgb}{0,.5,0} \colorlet{myPurple}{blue!40!red} \definecolor{myOrange}{rgb}{1,0.5,0}

\newcommand{\be}{\begin{equation}}\newcommand{\ee}{\end{equation}}
\newcommand{\bea}{\begin{eqnarray}}\newcommand{\eea}{\end{eqnarray}}
\newcommand{\nn}{\nonumber}
\def\({\left(}
\def\){\right)}
\def\[{\left[}
\def\]{\right]}

\newcommand{\si}{\sigma}

\newcommand{\Ga}{\Gamma}

\def\sD{\slashed{D}}

\renewcommand{\phi}{\varphi}

\begin{document}

\title{
Chiral magnetic effect at finite temperature in a field-theoretic approach
}

\author{C. G. Beneventano}
\email{gabriela@fisica.unlp.edu.ar}
\affiliation{Departamento de F\'isica and Facultad de Ingenier\'ia, Universidad Nacional de La Plata,
Instituto de F\'isica de La Plata,  CONICET--Universidad Nacional de La Plata,
C.C.67, 1900 La Plata, Argentina}%

\author{M. Nieto, E. M. Santangelo}
\email{mnieto@fisica.unlp.edu.ar, mariel@fisica.unlp.edu.ar}
\affiliation{Departamento de F\'isica, Universidad Nacional de La Plata,
C.C.67, 1900 La Plata, Argentina}%
\date{\today}
\begin{abstract}
We investigate the existence (or lack thereof) of the chiral magnetic effect in the framework of finite temperature field theory, studied through the path integral approach and regularized via the zeta function technique. We show that, independently of the temperature, gauge invariance implies the absence of the effect, a fact proved, at zero temperature and in a Hamiltonian approach, by N. Yamamoto. Indeed, the effect only appears when the manifold is finite in the direction of the magnetic field and gauge-invariance breaking boundary conditions are imposed. We present an explicit calculation for antiperiodic and periodic boundary conditions, which do allow for a CME, since only ``large'' gauge transformations are, then, an invariance of the theory. In both cases, the associated current does depend on the temperature, a well as on the size of the sample in the direction of the magnetic field, even for a temperature-independent chiral chemical potential. In particular, for antiperiodic boundary conditions, the value of this current only agrees with the result usually quoted in the literature on the subject in the zero-temperature limit, while it decreases with the temperature in a well-determined way.


\end{abstract}

\pacs{Valid PACS appear here}

\maketitle


\section{Introduction}

The relevance of the Dirac equation in the description of certain condensed matter systems, both two- and three-dimensional was recognized long time ago, as was the importance of Topology in giving such materials some exciting properties\cite{Nielsen:1983rb,Semenoff:1984dq,Haldane:1988zza}. After the production of graphene in the laboratory, the interest on these so-called Dirac materials received an ever-growing attention, both from a theoretical and an experimental point of view (for a review see, for instance, \cite{Wehling:2014cla}). Apart from graphene, Dirac matter includes topological insulators, Dirac and Weyl semimetals, among others.

The methods of Quantum Field Theory have proved very useful in the description of many aspects of graphene physics,(see, e.g., \cite{Gusynin:2007ix,Beneventano:2007fa,Vozmediano:2010zz,Fialkovsky:2011wh,Cortijo:2011aa,Beneventano:2013oma,Beneventano:2018tle}), as well as other Dirac materials \cite{Chernodub:2013kya,Witten:2015aoa,Pozo:2018yzs}. The aim of the present paper is to study, through such methods, the very interesting chiral magnetic effect (CME) \cite{Fukushima:2008xe,Kharzeev:2013ffa,Landsteiner:2016led}, i.e., the appearance of an electric current in the direction of the applied magnetic field, due to a chiral imbalance originated in the 4-d axial anomaly \cite{Adler:1969gk,Bell:1969ts} when a Dirac semimetal is placed in parallel electric and magnetic fields. Quite recently, the magnetic conductivity was measured in several three-dimensional Dirac/Weyl semimetals \cite{Kim:2013dia,Li:2014bha,d8fb42473715479798def7459a1a0553,Li_2015}.

To the best of our knowledge, this is the first study of the CME in the path integral approach to finite temperature field theory. Our complete calculation, which uses the gauge-invariant zeta function regularization \cite{Dowker:1975tf}, leads to some conclusions about the effect of the temperature and sample size which may be amenable to experimental test.

The outlay of the paper is the following: in section \ref{Sec2} we introduce our main conventions and determine the spectrum of the Euclidean four-dimensional Dirac operator in the presence of a constant chiral chemical potential and of a constant magnetic field. In particular, we study the properties of those modes (called ``special'' in this article, as opposed to those we call ``ordinary'' ones) which, as we show in section \ref{Sec3}, are responsible for the appearance of the CME whenever it is present.

Section \ref{Sec3} contains the calculation of the zeta function-regularized effective action, with a brief introduction to this well-known gauge-invariant regularization method. We first discuss the case of a continuous momentum, $k$, in the direction of the magnetic field, and argue that, as a consequence of gauge invariance, the CME does not exist in this case, whether at zero or finite temperature, a fact already conjectured in \cite{Valgushev:2015pjn} and proved, in a Hamiltonian approach and at zero temperature,  in \cite{PhysRevD.92.085011}. In the case of a manifold compactified in that direction with antiperiodic boundary conditions, we show that the ordinary modes do not contribute to the effect, and obtain the contribution of the special modes in a form which is adequate for taking the zero-temperature limit, where it coincides with the usually quoted result. We find that, by virtue of the invariance under ``large'' gauge transformations (see, for instance, \cite{dunne1999aspects} and references therein) both of the theory and of the regularization method, the current shows a periodic behavior as a function of the chiral chemical potential. In the same section, we also show that such current is by no means independent of the temperature. Indeed, the effect disappears at high temperatures (the inverse temperature $\beta$ much smaller than the length of the sample in the direction of the magnetic field), as is confirmed through an alternative expression of the effective action, obtained in appendix \ref{appendix}.

Finally, section \ref{Sec4} contains some comments and conclusions. Apart from summarizing our main results we also discuss, in this section, the case of periodic nonlocal boundary conditions, as well as the case of local, elliptic, gauge-invariant ones in the family of chiral bag boundary conditions \cite{Gilkey:2005qm,Beneventano:2003hv,Fialkovsky_2019}. In the case of local confining boundary conditions, our conclusions agree with previous results for finite size samples \cite{Gorbar_2015,PhysRevD.94.085014}.

Throughout the paper, we use natural units ($\hbar = c = k_B =1 $).

\section{Conventions and discussion of the spectrum}\label{Sec2}

\subsection{Determination of the spectrum} \label{spectrum}

We consider the Dirac equation in 4-d Euclidean space, in the presence of a constant positive magnetic field in the $x^3$ direction, as well as a  given chiral chemical potential ${\mu}_5$, which enforces the chiral imbalance. We also introduce a constant gauge field $\alpha$ in the direction of the magnetic field, to allow for the evaluation of the current $J^3$ by performing the $\alpha$-derivative of the effective action and evaluating it at $\alpha=0$.

Our Euclidean gamma matrices are given by
\bea
\Gamma^0 &=&\( \begin{array}{cc} 0 & I \\ I & 0 \end{array}\),\quad
\Gamma^i =\( \begin{array}{cc} 0 & i\si_i \\ -i \si_i & 0 \end{array}\),\, i=1,2,3,\quad \Gamma^5 =\( \begin{array}{cc} I & 0 \\ 0 & -I\end{array}\)
\,.\label{Gamma}
\eea

The Dirac operator to be used in order to obtain the finite-temperature effective action is
\bea
\sD=i\Ga^0\partial_0+i\Ga^0\Ga^5\mu_5+i\Ga^1\partial_1+\Ga^2(i\partial_2+eBx^1)+\Ga^3(i\partial_3+\alpha)\,,  \label{Dirac}
\eea
where $e$ is the absolute value of the electron charge and, as said, $B>0$.

Imposing antiperiodic conditions in the ``time'' direction in order to obtain the adequate Matsubara frequencies $\omega_l=(2l+1)\frac{\pi}{\beta}$, with $\beta$ the inverse temperature, and Fourier-transforming in $x^2$ and $x^3$,
\bea
\Psi(x^0,\vec{x})=e^{-i\omega_lx^0}e^{-ik_3x^3}e^{-ik_2x^2}\Psi(x^1),\quad l=-\infty,\ldots ,\infty\,.
\eea

The eigenvalue problem then reads
\bea
\sD\Psi=\left(\Ga^0\omega_l+i\Ga^0\Ga^5\mu_5+i\Ga^1\partial_1+\Ga^2(k_2+eBx^1)+\Ga^3(k_3+\alpha)\right)\Psi
=\lambda \Psi\,.
\eea

Now, defining the new variable $\xi=x^1+\frac{k_2}{eB}$, the previous equation can be rewritten as
\bea
\sD\Psi=\left(\Ga^0\omega_l+i\Ga^0\Ga^5\mu_5+i\Ga^1\partial_{\xi}+\Ga^2\,eB\,\xi+\Ga^3(k_3+\alpha)\right)\Psi=\lambda \Psi\,
\eea
or, in a more explicit form,
\bea
\( \begin{array}{cc} 0 & \omega_l-i \mu_5-\si_1\partial_{\xi}+i\si_2 eB\xi+i\si_3(k_3+\alpha) \\ \omega_l+i \mu_5+\si_1\partial_{\xi}-i\si_2 eB\xi-i\si_3(k_3+\alpha) & 0 \end{array}\) \(\begin{array}{c}
                                              \psi_1 \\
                                               \psi_2
                                             \end{array}\)=\lambda \(\begin{array}{c}
                                              \psi_1 \\
                                               \psi_2
                                             \end{array}\)\,.
\eea

So, we have
\be
\left(\omega_l-i \mu_5-\si_1\partial_{\xi}+i\si_2 eB\xi+i\si_3(k_3+\alpha)\right) \psi_2=\lambda \psi_1\label{psi1psi2}
\ee
and
\be
\left(\omega_l+i \mu_5+\si_1\partial_{\xi}-i\si_2 eB\xi- i\si_3(k_3+\alpha)\right) \psi_1=\lambda \psi_2\,.\label{psi2psi1}
\ee

It is easy to show that there is no solution for $\lambda=0$. So, after solving \eqref{psi2psi1} for $\psi_2$, and replacing into \eqref{psi1psi2}, we get
\be
\left({\omega_l}^2+(\mu_5-i\si_1\partial_{\xi}-\si_2 eB\xi-\si_3(k_3+\alpha))^2\right)\psi_1={\lambda}^2 \psi_1\,.
\ee

It can be seen, in a similar way, that $\psi_2$ satisfies the same equation.

There are two types of solutions to this problem,

\begin{itemize}

\item {Special modes}
\be
\Psi_{sp}(\xi)=C\,e^{-\frac{eB{\xi}^2}{2}}\(\begin{array}{c}
                                       \lambda \\
                                         0\\
                                        (\omega_l+i\Lambda)\\
                                        0
                                      \end{array}\)\,,\label{correntosos}\ee
which correspond to $\lambda= \pm \sqrt{{\omega_l}^2+{\Lambda}^2}$, with $\Lambda=\mu_5 - (k_3+\alpha)$, and $C$ the normalization factor. As we will show later, it is these modes that are responsible for the CME when the $x^3$ direction is compactified. In what follows, we will refer to these modes as the special ones. Note the correspondence between these eigenmodes of the Dirac operator, at zero temperature and for ${\mu}_5= \alpha=0$, and the lowest Landau modes of the Hamiltonian.

\item {Ordinary modes}

\be
\Psi_{ord}(\xi)=D\,e^{-\frac{eB{\xi}^2}{2}}\(\begin{array}{c}
                                       \lambda \, H_n(\xi) \\
                                         \frac{ -2ineB\lambda }{\Lambda_{\pm}-({\mu}_5+k_3+\alpha)}H_{n-1}(\xi)\\
                                        (\omega_l+i\Lambda_{\pm})H_n(\xi) \\
                                         \frac{ -(\omega_l+i\Lambda_{\pm})2ineB}{\Lambda_{\pm}-({\mu}_5+k_3+\alpha)} H_{n-1}(\xi)
                                      \end{array}\)\,.\label{comunes}\ee
Again, $\lambda= \pm \sqrt{{\omega_l}^2+{\Lambda_{\pm}}^2}$, $D$ is the normalization factor and $\Lambda_{\pm}=\mu_5 \pm\sqrt{2neB+(k_3 +\alpha)^2}$. We will call these modes ordinary ones.

\end{itemize}

In all cases, the usual Landau degeneracy $\frac{eB}{2\pi}$ is to be taken into account. In the case of a continuous $k_3$, the density of states $\frac{L_z}{2\pi}$ must also be used in order to obtain an effective action per unit area perpendicular to the magnetic field.

\subsection{Properties of the special modes}

Note that the special modes in equation \eqref{correntosos} are eigenfunctions of $i\Ga^0\Ga^3\Ga^5$  with eigenvalue $+1$ (those corresponding to the eigenvalue $-1$ fail to be square-integrable). As a consequence, they are zero modes of
$ i\Ga^1\partial_{\xi}+\Ga^2\,eB\,\xi \,.$

Equivalently, they satisfy $i\Ga^2\(\partial_{\xi}+\,eB\,\xi\)\Psi_{sp}=0$, which implies $\(\partial_{\xi}+\,eB\,\xi\)\Psi_{sp}=0$. So, the Landau degeneracy multiplied by the area is, in this case, nothing but the index of the operator $\(\partial_{\xi}+\,eB\,\xi\)$, which is not self adjoint, with its domain defined by the square-integrability condition.

Moreover, when considering these modes, the eigenvalue equation for the operator in \eqref{Dirac} reduces to
\bea
\left(i\Ga^0\partial _0+i\Ga^0\Ga^5\mu_5+\Ga^3(i\partial _3+\alpha)\right)\Psi = \left(i\Ga^0\partial _0+\Ga^3(i\partial _3+\alpha-\mu_5)\right)\Psi = \lambda \Psi\,. \label{gaugeinv}
\eea
This explains why, the corresponding eigenvalues depend on $\alpha -\mu_5$. It also shows that, at least part of this dependence, can be eliminated through a gauge transformation. We will analyze this point in the next section.

Also, by taking into account that $\Psi_{sp}(\xi)=({\varphi}_1,0,{\varphi}_2,0)^T$, it is evident that, as already stressed in \cite{Basar:2012gm}, the problem restricted to these modes is a $2$-d Euclidean one, i.e.,
\be
\left[{\sigma}_1 i{\partial}_0-{\sigma}_2(i{\partial}_3 +\alpha-\mu_5)\right]\left(
                                                                                \begin{array}{c}
                                                                                  {\varphi}_1 \\
                                                                                  {\varphi}_2\\
                                                                                \end{array}
                                                                              \right)=\lambda \left(
                                                                                \begin{array}{c}
                                                                                  {\varphi}_1 \\
                                                                                  {\varphi}_2\\
                                                                                \end{array}
                                                                              \right)\,.
\ee

\section{Effective action and current via zeta regularization} \label{Sec3}

In the framework of the path integral, the finite temperature effective action can be evaluated, by using the zeta-function regularization \cite{Dowker:1975tf}, as
\be S_{\rm eff}=-\log { Z}_\beta\equiv\left.\frac{d}{ds}\right\vert_{s=0}\zeta\(\frac{\sD}{\rho},s\)\,,\label{seff}\ee
where
$$\zeta\(\frac{\sD}{\rho},s\)=\sum_{\lambda} \(\frac{\lambda}{\rho}\)^{-s}\,, \label{zeta}$$
and $\rho$ is a parameter with dimension of mass, introduced to render the argument of the zeta function dimensionless.

The current in the direction of the magnetic field, per unit area perpendicular to it, will then be given by
\be\left.J^3= -\frac{e}{\beta}\frac{\partial}{\partial \alpha}\right\vert_{\alpha=0}S_{\rm eff}\,.
\label{J}
\ee

\subsection{Case of a continuous $k_3$. Absence of CME for an unbounded region of space}

We will first consider the unbounded-space case. The zeta function can be written as
$$\zeta\(\frac{\sD}{\rho},s\)= \zeta _{sp} (s) + \zeta _{ord} (s), $$
where $\zeta _{sp} (s)$ comes from the modes in equation \eqref{correntosos}, while $\zeta _{ord} (s)$ comes from the modes in equation \eqref{comunes}.

From the explicit expressions of both types of eigenvalues, it is evident that, once zeta-regularized, and as a consequence of gauge invariance, there is no current parallel to the magnetic field when $x^3$ is unbounded and, thus, $k_3$ is continuous. Indeed, $\alpha$ can be trivially absorbed into a shift of $k_3$, thus leading to an $\alpha$-independent effective action and, consequently, to no CME. As an example of how gauge invariance works, we evaluate the contribution of the special modes to the effective action in appendix \ref{appendix1}. Note this is in disagreement with the unrenormalized result obtained, for instance, in \cite{Kharzeev:2013ffa} and extends to finite temperature the result of \cite{PhysRevD.92.085011}.

\subsection{Case of a discrete $k_3$. Existence of CME for antiperiodic boundary conditions}

Now, we turn to the case of a compactified $x^3$ direction. As an example, we will impose on the eigenfunctions antiperiodic boundary conditions , i.e. $k_3$ takes discrete values given by $q_k=(2k+1)\frac{\pi}{L_z}$, so the eigenfunctions can be written as
\bea
\Psi(x^0,\vec{x})=e^{-i\omega_lx^0}e^{-iq_k x^3}e^{-ik_2x^2}\Psi(x^1)\quad l,k=-\infty,\ldots ,\infty\,.
\eea

\subsubsection{Contribution of the ordinary modes}

Recalling the definition of $\Lambda _{\pm}$, which, in the case of these boundary conditions reduces to $\Lambda _{\pm}=\mu_5 \pm\sqrt{2neB+(q_k +\alpha)^2}$, the contribution of the ordinary modes \eqref{comunes} to the zeta function is given by
\bea
\zeta _{ord} (s) &=& \(1 + (-1)^{- \frac{s}{2}}\)\frac{eB}{2\pi} \(\frac{1}{\rho}\)^{-s}  \!\!\!\sum _{k,l=-\infty}^{\infty} \sum _{n=1}^{\infty} \left\{ \[\omega_l^2 + \Lambda_+^2\]^{-\frac{s}{2}} + \[\omega_l^2 + \Lambda_-^2\]^{-\frac{s}{2}} \right\}\nn\\
&=& \frac{\(1 + (-1)^{- \frac{s}{2}}\)eB}{2\pi}\(\frac{1}{\rho}\)^{-s}  \!\!\!\sum _{l=-\infty}^{\infty} \sum _{n=1}^{\infty}\sum _{k=0}^{\infty} \left\{ \[\omega_l^2 + \(\mu_5+\sqrt{2neB+\(\alpha + (2k+1)\frac{\pi}{L_z}\)^2}\)^2\]^{-\frac{s}{2}} \right.\nn\\
&+& \left.\[\omega_l^2 +\(\mu_5-\sqrt{2neB+\(\alpha + (2k+1)\frac{\pi}{L_z}\)^2}\)^2\]^{-\frac{s}{2}} \right\}+\left\{\alpha \rightarrow -\alpha\right\}
\,.
\eea
From the last expression, it is evident that the part of the effective action coming from these modes will be even in $\alpha$, thus giving no contribution to $J^3$, no matter the values of $\mu_5$, $B$ or the temperature.

\subsubsection{Contribution of the special modes}

In this case, the contribution to the zeta function becomes
\bea
\zeta _{sp}(s) &=& \(1 + (-1)^{- \frac{s}{2}}\)\frac{eB}{2\pi} \rho^s \sum _{l=-\infty}^{\infty} \sum _{k=-\infty}^{\infty} \frac{1}{\Gamma(s/2)} \int _{0}^{\infty} dt\, t^{\frac{s}{2}-1} e^{- \[\omega_l^2 + \(q_k + \alpha - \mu_5\)^2\]t}\,. \label{zsp}
\eea

In order to isolate the zero-temperature ($\beta \rightarrow \infty$) limit, we use the inversion formula for the Jacobi Theta function in the index $l$, which gives as a result
\bea
\zeta _{sp} (s) &=& \(1 + (-1)^{- \frac{s}{2}}\)\frac{eB}{2\pi} \rho^s {\(\frac{2\pi}{\beta}\)}^{-s}\frac{{\pi}^{\frac12}}{\Gamma(s/2)}\sum _{l=-\infty}^{\infty} \sum _{k=0}^{\infty}  \int _{0}^{\infty} dt\, t^{\frac{s-1}{2}-1}(-1)^l e^{- \frac{{\pi}^2 l^2}{t}} \[e^{-\([q_k+ \alpha - \mu_5]\frac{\beta}{2\pi}\)^2t}\right.\nonumber\\
&+&\left.e^{-\([q_k+ \mu_5- \alpha]\frac{\beta}{2\pi}\)^2t}\]\nn \\
&=& \(1 + (-1)^{- \frac{s}{2}}\)\frac{eB}{2\pi} \rho^s {\(\frac{2\pi}{\beta}\)}^{-s}\frac{{\pi}^{\frac12}}{\Gamma(s/2)}\left\{\sum _{k=0}^{\infty} \Gamma\(\frac{s-1}{2}\) \(\frac{\beta}{2\pi}\)^{1-s} \left[\(q_k+ \alpha - \mu_5\)^2\right]^{\frac{1-s}{2}} \right. \nn \\
&+& \left. 2 \sum _{l=1}^{\infty} \sum _{k=0}^{\infty}  \int _{0}^{\infty} dt\, t^{\frac{s-1}{2}-1}(-1)^l e^{- \frac{{\pi}^2 l^2}{t}} e^{-\([q_k+ \alpha - \mu_5]\frac{\beta}{2\pi}\)^2t} \right\} + \left\{\(\alpha - \mu_5 \)\rightarrow \( \mu_5 - \alpha  \)\right\}
\,. \label{z1}
\eea

After performing the $t$-integral, and making use of the definition of the Hurwitz zeta function $\zeta_H\(s,q\)=\sum _{k=0}^{\infty}(k+q)^{-s}$, valid for $\Re{s}>1$ and $q>0$ \cite{zwillinger2014table}
\bea
\zeta _{sp} (s)&=& \(1 + (-1)^{- \frac{s}{2}}\)\frac{eB}{2\pi}\(\frac{\beta}{2\pi}\)^s \rho^s \frac{s{\pi}^{\frac12}}{2\Gamma(s/2+1)}\[\Gamma\(\frac{s-1}{2}\){\(\frac{\beta}{L_z}\)}^{1-s}\zeta_H\(s-1,\frac12+(\alpha-\mu_5)\frac{L_z}{2\pi}\)\right.\nonumber  \\
&+&\left. 4 \sum _{l=1}^{\infty} \sum _{k=0}^{\infty} (-1)^l  \(\frac{2{\pi}^2 l }{\beta (q_k+\alpha-\mu_5)}\)^{\frac{s-1}{2}} K_{\frac{s-1}{2}}\(l\beta(q_k+\alpha-\mu_5)\)\]\nonumber  \\
&+& \left[\(\alpha - \mu_5 \)\rightarrow \( \mu_5 - \alpha  \)\right]\,,\label{zesp}
\eea
where $K_{\nu} (x)$ is the modified Bessel function of order $\nu$ as defined, for instance, in \cite{zwillinger2014table}. Note that this result is valid for $|\mu _5-\alpha|<\frac{\pi}{L_z}$. We will comment on this point later in this section.

As in the unbounded case, $\zeta _{sp} (0)=0$, which makes the evaluation of the effective action direct, and gives a result which is independent from the unphysical parameter $\rho$. The special contribution to the effective action is, always for $|\mu _5-\alpha|<\frac{\pi}{L_z}$, given by
\bea
S_{{\rm eff},sp}&=&\frac{eB}{2\pi} \left\{\frac{2 \pi \beta}{L_z} \frac{1}{2}\left( -\frac{1}{12} + \(\frac{L_z (\alpha - \mu_5)}{2 \pi}\)^2\right)-2 \sum _{k=0}^{\infty} \log{\(1+e^{-\beta \(q_k+\alpha-\mu_5\)}\)}\right\}\nonumber \\
&+&\left\{\(\alpha - \mu_5 \)\rightarrow \( \mu_5 - \alpha  \)\right\}\,.
\label{seffsp}
\eea

The zero-temperature contribution to the effective action and to the current can be retrieved from the first term in the last equation which corresponds to $l=0$ in equation \eqref{zesp}. The remaining values of $l$, instead, will give contributions decaying exponentially with the inverse temperature, $\beta$.


Here, it is interesting to note that, due to the particular properties of the special modes already discussed, once the large gauge transformation is performed for any temperature, in the zero-temperature limit, where the compactification of the ``time'' coordinate becomes irrelevant, the remaining part of $\alpha - \mu_5$ can be classically eliminated through a chiral transformation in $x^0$. In this limit, the term depending on  $\alpha - \mu_5$ is nothing but the mass term expected from the chiral anomaly in $1+1$ dimension (see, for instance, equation (2.35) in reference \cite{GamboaSaravi:1981zd}). This fact has already been stressed in reference \cite{Basar:2012gm}.

According to equation \eqref{J}, the current in the direction of the magnetic field per unit area is, in the zero-temperature limit,
\be
J^{3 (l=0)}_{sp} = \frac{e^2B L_z}{2\pi^2} \mu_5 \,\label{J10}.
\ee

Once this expression is multiplied by the area, equation (24) in \cite{Kharzeev:2013ffa} is obtained. However, we stress two important issues: In the first place, \eqref{J10} is obtained only in the case a compactified $x^3$ coordinate and for some nonlocal boundary conditions. In the second place, while it is true that this is the only contribution to the chiral magnetic current in the zero-temperature limit ($\beta \rightarrow \infty$), at any finite temperature, there are exponential corrections coming from this eigenspace for $l\neq 0$, given (always in the range $|\alpha - \mu _5|<\frac{\pi}{L_z}$) by
\bea
J^{3 (l\neq 0)}_{sp}= -\frac{e^2B }{\pi}\sum_{n=0}^{\infty}\[\frac{e^{-\beta\left(q_n-\mu_5\right)}}{1+e^{-\beta\left(q_n-\mu_5\right)}}-
\frac{e^{-\beta\left(q_n+\mu_5\right)}}{1+e^{-\beta\left(q_n+\mu_5\right)}}\]\,. \label{J1no0}
\eea

\begin{figure}[h!]
\centerline{ \includegraphics[height=4.truecm]{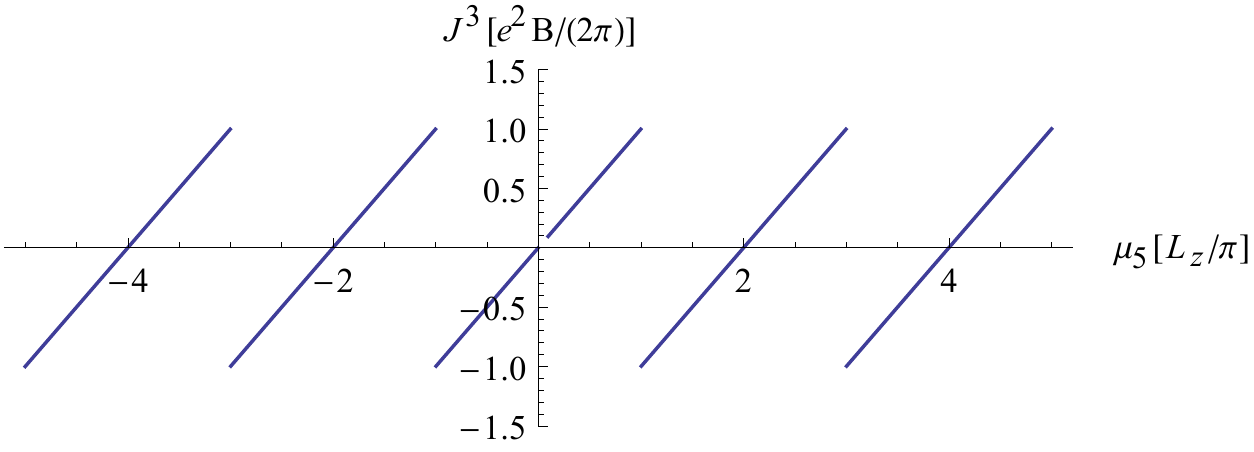} }
\caption{Current density at zero temperature as a function of the chiral chemical potential.}\label{figure1}
\end{figure}

Both \eqref{J10} and \eqref{J1no0} are valid for $|\mu _5|<\frac{\pi}{L_z}$. It is enough to study this range, since the periodicity of the result is a consequence of the fact that, this time, only ``large'' gauge transformations preserve the anti-periodicity imposed when compactifying the $x^3$ direction \cite{sissakian1998topological}. In fact, for these special modes, according to equation \eqref{gaugeinv}, one can always transform $\Psi \rightarrow e^{i \frac{2 \pi}{L_z} n x^3} \Psi$, with $n \in \mathbb{Z}$, that does not spoil de antiperiodic boundary condition imposed on $x^3$, which leads to $\alpha - \mu_5 \rightarrow \alpha - \mu_5-  \frac{2 n \pi}{L_z}$. The zero-temperature part of the chiral magnetic current is shown in FIG. \ref{figure1}

\begin{figure}[h!]
\centerline{ \includegraphics[height=9.truecm]{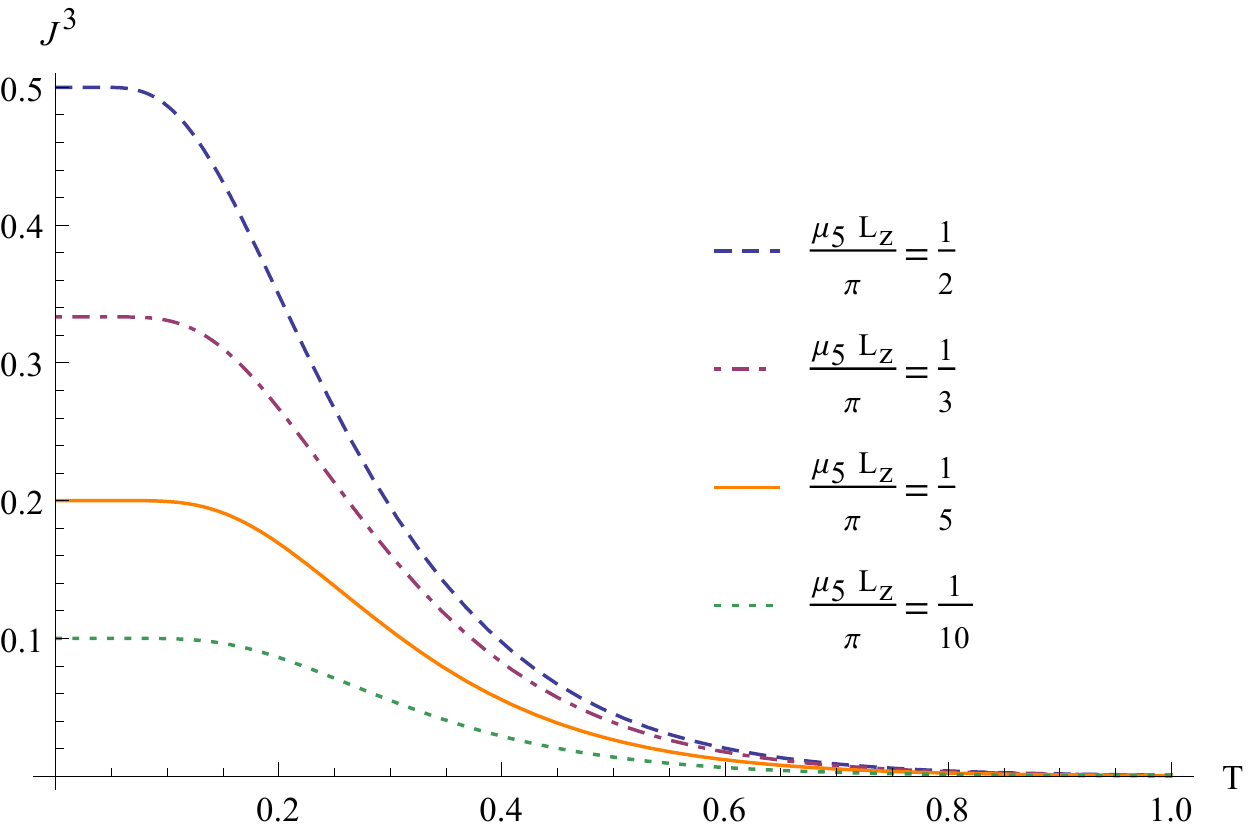} }
\caption{Current density (in units of $\frac{e^2B}{2\pi}$) as a function of the temperature (in units of $\frac{L_z}{\pi}$), for some values of the axial chemical potential.}\label{figure2}
\end{figure}

We stress that, contrarily to what is sometimes stated, the current associated to the CME is, by no means, temperature-independent. The dependence of $J^3$ on the temperature, for different values of $\mu_5$ in the range $|\mu _5|<\frac{\pi}{L_z}$ is depicted in FIG. \ref{figure2}.






\section{Conclusions}\label{Sec4}

To summarize, we have performed a full computation of the current associated to the chiral magnetic effect, in the framework of finite temperature field theory, regularized through the gauge-invariant zeta function method in the path integral formulation. From our calculation, some remarkable conclusions arise:

In the first place, there is no CME when the manifold is infinite in the $x^3$-direction, which is an evident consequence of gauge invariance, preserved by our regularization method. From our detailed calculations it is evident that any other gauge-invariant regularization will lead to the same conclusion, which is consistent with the results found, for instance, in reference \cite{Hou_2011}, where canonical quantization was performed, together with a Pauli-Villars regularization, and also in \cite{Rebhan_2010}, where an holographic approach was employed. Moreover, a generalization of Bloch's no-go theorem was, indeed, shown to hold for relativistic fields, on the basis of gauge invariance \cite{PhysRevD.92.085011}.

On the other hand, we have shown that, in the compact case with antiperiodic boundary conditions, the CME arises from the special modes, whose properties we studied in detail. The associated current coincides, at zero temperature, with the result usually quoted for the non-compact case (see, for instance, \cite{Kharzeev:2013ffa}), and seemingly considered to be valid at any temperature in the case of a constant chiral potential. Here, always as a consequence of gauge invariance, this time meaning invariance under ``large" gauge transformations, we found that the CME current depends on $\mu_5$ in a periodic way. This dependence appears in FIG. \ref{figure1}. We understand this point would be something worth exploring experimentally. As is well known, the zero-temperature expression, combined with the assumption that the chiral potential $\mu_5$ is proportional to the four-dimensional anomaly gives, for parallel electric and magnetic fields, a longitudinal magnetic conductance proportional to the square of the magnetic field, something that seems to have been confirmed experimentally by now \cite{Kim:2013dia,Li:2014bha,d8fb42473715479798def7459a1a0553,Li_2015}.

Moreover, our calculation shows that, for a given length $L_z$, the CME current decreases as the temperature increases. In this article, we have determined its precise dependence on the temperature. In particular, we have reobtained the disappearance of the effect in the noncompact case, by explicitly performing the infinite temperature limit in appendix \ref{appendix}. The disappearance of the effect with the temperature is usually attributed to a dependence of the chiral chemical potential with the square of the inverse temperature. We find, instead, that the effect decreases with the temperature, even when $\mu_5$ is a constant. Here, we have presented a clear deduction of the way it decreases. The temperature dependence for given values of the remaining variables appears in FIG. \ref{figure2}. We understand that it would also be interesting to contrast this point with experiments.

Now, our explicit calculation leads to some predictions for other boundary conditions, both nonlocal and local ones, in a finite size sample. In the first place, it is easy to see that, in the case of periodic (instead of antiperiodic) nonlocal boundary conditions, the results are essentially the same as in our case, except for a shift in the dependence on $\mu_5$. As concerns local elliptic boundary conditions of the bag or even chiral bag type \cite{Beneventano:2003hv}, there will not be a CME at all, since such boundary conditions are gauge-invariant. This is in total agreement with previous results as, for instance, the ones obtained in reference \cite{Gorbar_2015} for a slab with pure bag boundary conditions. Indeed, in this reference, the CME for a constant chiral chemical potential was shown to vanish, not only in mean but also locally. The same conclusion was arrived to in reference \cite{PhysRevD.94.085014}, which presents a generalization of the results in \cite{Gorbar_2015} to the most general confining local boundary conditions and to finite temperature.

\appendix

\section{Contribution of the special modes to the effective action in the unbounded case}\label{appendix1}

As an example of how gauge invariance prevents the existence of CME at any temperature, we evaluate here the zeta-regularized contribution of the special modes to the effective action and show how, once the gauge invariant zeta regularization is applied, the result is $\alpha$-independent. For these modes,
\bea
\zeta _{sp} (s) &=& \(1 + (-1)^{- \frac{s}{2}}\)\frac{eB}{2\pi} \frac{L_z}{2\pi} \rho^s \sum _{l=-\infty}^{\infty} \int _{-\infty}^{\infty} dk_3 \[ \omega_l^2 + \( k_3 + \alpha - \mu_5\)^2\]^{-\frac{s}{2}}\,.
\eea

In order to perform the analytic extension of $\zeta_{sp}$, we Mellin-transform this last expression, and get
\bea
\zeta _{sp} (s) &=& \(1 + (-1)^{- \frac{s}{2}}\)\frac{eB}{2\pi} \frac{L_z}{2\pi} \rho^s \sum _{l=-\infty}^{\infty} \int _{-\infty}^{\infty} dk_3 \frac{1}{\Gamma(s/2)} \int _{0}^{\infty} dt\, t^{\frac{s}{2}-1} e^{- \[\omega_l^2 + \( k_3 + \alpha - \mu_5\)^2\]t}\,,
\eea
which, after changing variable to $\kappa = k_3 + \alpha$, and performing the $\kappa$ integration, leads to
\bea
\zeta _{sp} (s) &=& \(1 + (-1)^{- \frac{s}{2}}\)\frac{eB}{2\pi} \frac{L_z}{2\pi} \rho^s \sum _{l=-\infty}^{\infty} \frac{\sqrt{\pi}}{\Gamma(s/2)} \int _{0}^{\infty} dt\, t^{\frac{s-1}{2}-1} e^{- \omega_l^2 t} \nn \\ \nn
&=& \(1 + (-1)^{- \frac{s}{2}}\)\frac{eB}{2\pi} \frac{L_z}{2\pi} \rho^s \sum _{l=-\infty}^{\infty} \frac{\sqrt{\pi}}{\Gamma(s/2)} \Gamma\(\frac{s-1}{2}\) \((2l +1)\frac{\pi}{\beta}\)^{1-s}\\ &=& \(1 + (-1)^{- \frac{s}{2}}\)\frac{eB}{2\pi} \frac{L_z}{2\pi} \rho^s \frac{s\sqrt{\pi}}{2 \Gamma(s/2+1)} \Gamma\(\frac{s-1}{2}\)\(\frac{2 \pi}{\beta}\)^{1-s} 2 \zeta_H(s-1,\frac{1}{2})\,.
\eea
Now, as can be easily seen, this part of the zeta function vanishes at $s=0$. So, it is easy to perform the $s$-derivative to get this partial contribution to the effective action, which we will call $S_{ {\rm eff}, sp}$
\bea
S_{{\rm eff}, sp}=-4\pi\frac{eB}{2\pi} \frac{L_z}{\beta} \zeta_H(-1,\frac{1}{2})\,. \label{noacotado}
\eea

It is clear that, this expression being $\alpha$-independent, gives no contribution whatsoever to a current in the direction of the magnetic field. This lack of CME is a consequence of the gauge invariance of the problem, which is well-known to be preserved by the zeta-function regularization. In fact, $\alpha - \mu _5$ can be removed from equation \eqref{gaugeinv} through a gauge transformation.

\section{High temperature limit of $J^3$}\label{appendix}

As is well known (see, for instance,\cite{Asorey:2012vp}), in order study the high temperature limit of the CME, i.e. the  $\beta  \rightarrow 0$-limit of $J^3$, we use the inversion formula of the Jacobi Theta function in the index $k$ in \eqref{zsp},
\bea
\zeta _{sp} (s) &=& \(1 + (-1)^{- \frac{s}{2}}\)\frac{eB}{2\pi} \rho^s \sum _{l=-\infty}^{\infty} \sum _{k=-\infty}^{\infty} \frac{1}{\Gamma(s/2)} \int _{0}^{\infty} dt\, t^{\frac{s}{2}-1} e^{- \[\omega_l^2 + \(q_k + \alpha - \mu_5\)^2\]t} \nn \\
 &=& \(1 + (-1)^{- \frac{s}{2}}\)\frac{eB}{2\pi} \rho^s {\(\frac{2\pi}{L_z}\)}^{-s}\frac{{\pi}^{\frac12}}{\Gamma(s/2)}\sum _{l, k =-\infty}^{\infty}  \int _{0}^{\infty} dt\, t^{\frac{s-1}{2}-1}(-1)^k e^{- \frac{{\pi}^2 k^2}{t}} e^{i\,k (\alpha - \mu_5) L_z} e^{-\left((2 l +1) \frac{L_z}{2 \beta}\right)^2 t}\nn \\
&=& \(1 + (-1)^{- \frac{s}{2}}\)\frac{eB}{2\pi} \rho^s {\(\frac{2\pi}{L_z}\)}^{-s}\frac{{\pi}^{\frac12}}{\Gamma(s/2)}\left\{\sum _{l=-\infty}^{\infty} \Gamma\(\frac{s-1}{2}\) \left[\((2l+1)\frac{L_z}{2 \beta}\)^2\right]^{\frac{1-s}{2}} \right. \nn \\
&+& \left. 2 \sum _{k=1}^{\infty} \sum _{l=-\infty}^{\infty}  \int _{0}^{\infty} dt\, t^{\frac{s-1}{2}-1}(-1)^k e^{- \frac{{\pi}^2 k^2}{t}} \cos(k (\alpha -\mu _5) L_z)\,e^{-\((2l+1)\frac{L_z}{2 \beta}\)^2t} \right\} \nn \\
&=& \(1 + (-1)^{- \frac{s}{2}}\)\frac{eB}{2\pi} \rho^s {\(\frac{2\pi}{L_z}\)}^{-s}\frac{{\pi}^{\frac12}}{\Gamma(s/2)}\left\{2 \Gamma\(\frac{s-1}{2}\) \(\frac{L_z}{\beta}\)^{1-s} \zeta_H\(s-1,\frac12\) \right. \nn \\
&+& \left. 8 \sum _{k=1}^{\infty} \sum _{l=0}^{\infty} (-1)^k \cos(k (\alpha -\mu _5) L_z) \(\frac{2\pi k \beta}{(2l+1)L_z}\)^{\frac{s-1}{2}} K_{\frac{s-1}{2}}\(\frac{\pi k (2l+1)L_z}{\beta}\)\right\}
\,.
\eea

Performing the $s$-derivative, and evaluating at $s=0$, we obtain an alternative expression for the effective action in equation \eqref{seffsp}, which allows for the high temperature limit to be taken.

\bea
S_{{\rm eff},sp}^{k=0}= -4\pi\frac{eB}{2\pi} \frac{L_z}{\beta} \zeta_H(-1,\frac{1}{2})\,.
\eea
This coincides with equation \eqref{noacotado}, this contribution to $S_{{\rm eff},sp}$ being exactly all that is obtained from the special modes in the unbounded space case. As stated before, it gives no contribution to the current in the direction of the magnetic field.

The $k\neq 0$ terms contribute to $S_{{\rm eff},sp}$ with
\bea
S_{{\rm eff},sp}^{k\neq 0}= 4 \frac{eB}{2\pi} \sum _{k=1}^{\infty} \sum _{l=0}^{\infty} \frac{(-1)^k}{k} \cos(k (\alpha -\mu _5) L_z)\,e^{- \frac{\pi k (2l+1) L_z}{\beta}}\,.
\eea

Given that, as already explained, the ordinary modes do not give any contribution to the current, all that is left comes from this last equation. We thus obtain an alternative expression for the current $J^3$, given by

\bea
J^3= -\frac{2 e^2B}{\pi \beta} \sum _{k=1}^{\infty} \sum _{l=0}^{\infty} (-1)^k L_z \sin(k \mu _5 L_z)\,e^{- \frac{\pi k (2l+1) L_z}{\beta}}\,.
\eea
From this expression, it is easily seen that the current in the direction of the magnetic field vanishes in the high temperature limit.

\section*{Acknowledgments}

We thank Mar\'{\i}a A.H. Vozmediano for several discussions and suggestions. We also thank the authors of reference \cite{Gorbar_2015} for calling our attention to their work and for some useful comments, as well as the referees for calling our attention to references \cite{PhysRevD.92.085011} and \cite{PhysRevD.94.085014}. This work was partially supported by CONICET (PIP 688) and UNLP (Proyecto Acreditado X-748).

\bibliography{CME7clean}
\bibliographystyle{apsrev4-1}

\end{document}